\documentclass{article}

\usepackage{CJKutf8}
\usepackage{longcat_style}
\usepackage{adjustbox}
\usepackage[utf8]{inputenc} %
\usepackage[T1]{fontenc}    %
\usepackage{newunicodechar}
\usepackage{hyperref}       %
\usepackage{xcolor}
\usepackage[normalem]{ulem} %
\hypersetup{
    colorlinks=true,      %
    linkcolor=blue,      %
    urlcolor=blue,       %
    citecolor=blue,      %
    linkbordercolor=blue, %
    urlbordercolor=blue,
    citebordercolor=blue,
    pdfborderstyle={/S/U/W 1}, %
}
\usepackage{float}
\usepackage{url}            %
\usepackage{booktabs}       %
\usepackage{amsfonts}       %
\usepackage{nicefrac}       %
\usepackage{microtype}      %
\usepackage{lipsum}		%
\usepackage{graphicx}
\usepackage{natbib}
\usepackage{doi}
\usepackage{amsmath}
\usepackage{amssymb} %
\usepackage{xspace}
\usepackage{enumitem}
\usepackage{multirow}
\usepackage{subcaption} 
\usepackage{makecell}
\usepackage{hyperref, cleveref}
\usepackage{pifont}
\usepackage[inkscapelatex=false]{svg}
\usepackage{caption}
\DeclareCaptionLabelSeparator{pipe}{ | }
\usepackage{tcolorbox}
\newtcolorbox{coloredquote}[1][]{
    colback=green!5!white,  
    colframe=green!70!black, 
    boxrule=2pt,
    arc=7pt,
    left=6pt,
    right=6pt,
    top=4pt,
    bottom=4pt,
    title=#1
}

\mathchardef\mhyphen="2D

\def\ttsname{LongCat-AudioDiT}

\title{\ttsname{}: High-Fidelity Diffusion Text-to-Speech in the Waveform Latent Space}

\author{
\begin{tabular}{c}
Meituan LongCat Team
\end{tabular}\\
\texttt{longcat-team@meituan.com} \\
}

\renewcommand{\headeright}{\raisebox{-0.2\height}{\includegraphics[height=1.8em]{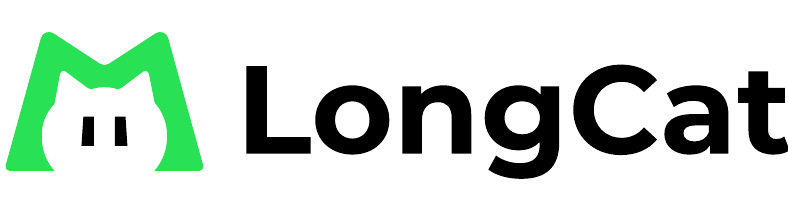}}}
\renewcommand{\shorttitle}{\ttsname{}}

\begin{document}
\maketitle
\setcounter{footnote}{0}

\begin{abstract}
We present \ttsname{}, a novel, non-autoregressive diffusion-based text-to-speech (TTS) model that achieves state-of-the-art (SOTA) performance.
Unlike previous methods that rely on intermediate acoustic representations such as mel-spectrograms, the core innovation of \ttsname{} lies in operating directly within the waveform latent space. This approach effectively mitigates compounding errors and drastically simplifies the TTS pipeline, requiring only a waveform variational autoencoder (Wav-VAE) and a diffusion backbone.
Furthermore, we introduce two critical improvements to the inference process: first, we identify and rectify a long-standing training-inference mismatch; second, we replace traditional classifier-free guidance with adaptive projection guidance to elevate generation quality.
Experimental results demonstrate that, despite the absence of complex multi-stage training pipelines or high-quality human-annotated datasets, \ttsname{} achieves SOTA zero-shot voice cloning performance on the Seed benchmark while maintaining competitive intelligibility.
Specifically, our largest variant, \ttsname{}-3.5B, outperforms the previous SOTA model (Seed-TTS), improving the speaker similarity (SIM) scores from 0.809 to 0.818 on Seed-ZH, and from 0.776 to 0.797 on Seed-Hard.
Finally, through comprehensive ablation studies and systematic analysis, we validate the effectiveness of our proposed modules.
Notably, we investigate the interplay between the Wav-VAE and the TTS backbone, revealing the counterintuitive finding that superior reconstruction fidelity in the Wav-VAE does not necessarily lead to better overall TTS performance.
Code and model weights are released to foster further research within the speech community.\\
\textbf{Github}:\url{https://github.com/meituan-longcat/LongCat-AudioDiT}\\
\textbf{HuggingFace}:\\
\url{https://huggingface.co/meituan-longcat/LongCat-AudioDiT-3.5B}\\
\url{https://huggingface.co/meituan-longcat/LongCat-AudioDiT-1B}
\end{abstract}
\section{Introduction}
Text-to-speech (TTS) synthesis is a fundamental task in content generation.
Recent TTS systems, built upon either autoregressive (AR) or non-autoregressive (NAR) generative paradigms, have achieved impressive speech quality that approaches human-level naturalness~\citep{wang2023valle, le2024voicebox, anastassiou2024seed, ju2024ns3, du2025cosyvoice3, zhang2025minimax}.
Among these paradigms, NAR TTS—particularly diffusion-based models—stands out for its generation quality, architectural simplicity, and inference efficiency.
Specifically, because NAR TTS can operate directly on continuous acoustic representations without relying on discrete audio tokenizers, it inherently bypasses complex system designs.
Although early NAR systems heavily relied on auxiliary duration prediction modules to establish temporal alignment between text and audio~\citep{ren2019fastspeech, le2024voicebox}, recent advances have demonstrated that models can implicitly learn this alignment given sufficient training data~\citep{eskimez2024e2, chen2024f5tts, lee2024ditto}, enabling further architectural simplification.
Furthermore, by generating the entire speech sequence in parallel, NAR TTS exhibits a distinct speed advantage over its AR counterparts, especially as the sequence length increases.
Despite these advantages, hybrid architectures that integrate both AR and NAR technologies have recently dominated the SOTA landscape~\citep{betker2023tortoise, anastassiou2024seed, du2024cosyvoice, zhang2025minimax}, generally outperforming pure diffusion-based NAR models~\citep{chen2024f5tts, lee2024ditto}.
An exception is the diffusion-based variant Seed-DiT, which reportedly surpasses its hybrid counterpart, Seed-ICL, within the Seed-TTS framework~\citep{anastassiou2024seed}.
However, the exact architecture and technical details of Seed-DiT remain undisclosed, leaving a critical gap regarding how to construct a pure, highly performant diffusion-based TTS system.

In this paper, we present \ttsname{}, a diffusion-based NAR TTS model that achieves SOTA performance.
A core finding of our work is that training the diffusion model directly in the waveform latent space yields substantial improvements over traditional paradigms that rely on intermediate acoustic representations, such as mel-spectrograms.
Consequently, \ttsname{} consists of only two streamlined components: a waveform variational autoencoder (Wav-VAE)~\citep{kingma2013vae} and a diffusion Transformer (DiT)~\citep{vaswani2017attention, peebles2023dit}.
During training, the VAE encoder produces continuous latents for the DiT.
During inference, the VAE decoder synthesizes raw waveforms directly from the latents sampled by the DiT, completely bypassing intermediate representations and eliminating the need for auxiliary vocoders heavily relied upon in previous studies~\citep{chen2024f5tts, lee2024ditto}.
This end-to-end design mitigates the compounding errors typically incurred when predicting mel-spectrograms and subsequently converting them into waveforms.
To support robust multilingual synthesis, we condition the model not only on the last hidden states but also on the raw word embeddings extracted from a pretrained language model.
Furthermore, we introduce two critical improvements to the inference process: first, we identify and rectify a long-standing training-inference mismatch; second, we replace traditional classifier-free guidance with adaptive projection guidance to elevate generation quality.
Finally, we explore the scalability of our architecture and observe a clear performance advantage when scaling up the model size.
The final version of \ttsname{}, comprising 3.5B parameters and trained on 1 million hours of Chinese and English speech data, achieves SOTA performance on the Seed benchmark~\citep{anastassiou2024seed}.
To thoroughly validate our approach, we conduct comprehensive ablation studies on the proposed techniques.
In addition, we systematically investigate the impact of latent dimensionality and compression rates on both the reconstruction fidelity of the Wav-VAE and the overall generation quality of the TTS model.

\begin{figure}[t]
\centering
\setlength{\belowcaptionskip}{-.5cm}
\includegraphics[width=0.8\columnwidth]{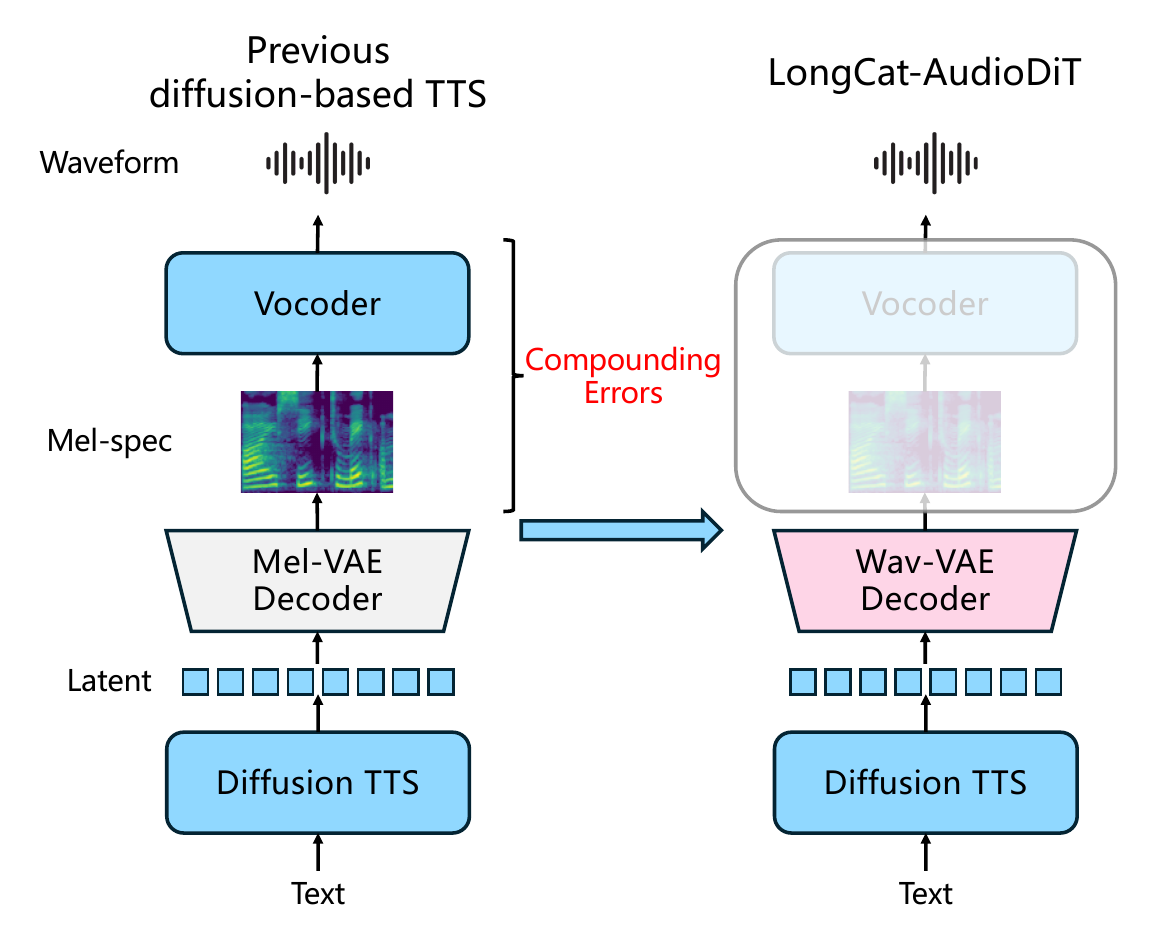}
\caption{Overview of \ttsname{}. Our architecture generates continuous waveform latents directly, thereby avoiding the compounding errors that inherently arise when predicting and subsequently converting intermediate representations (e.g., mel-spectrograms) into waveforms.}
\label{figure:overview}
\end{figure}

Our main contributions are summarized as follows:
\begin{itemize}
    \item We propose \ttsname{}, a SOTA diffusion-based NAR TTS model. By operating directly in the waveform latent space, our approach effectively eliminates the compounding errors introduced by intermediate representations like mel-spectrograms.
    \item We propose two critical improvements to the inference process: first, we identify and rectify a long-standing training-inference mismatch; second, we replace traditional classifier-free guidance with adaptive projection guidance to elevate generation quality.
    \item We conduct systematic and comprehensive experiments to validate the effectiveness of our design choices. Notably, we provide empirical insights into the non-trivial relationship between the reconstruction quality of the Wav-VAE and the ultimate synthesis quality of the TTS backbone.
    \item We publicly release the source code and model weights of \ttsname{} to advance research and development within the community.
\end{itemize}

\section{Related Work}
\subsection{Diffusion-based TTS}
Early diffusion-based TTS models, such as Grad-TTS~\citep{popov2021gradtts} and Diff-TTS~\citep{jeong2021difftts}, adopted diffusion probabilistic models (DPMs)~\citep{sohl2015dpm, song2020score, ho2020ddpm} governed by stochastic differential equations (SDEs).
The fundamental concept of these approaches is to construct a bidirectional transformation between a simple Gaussian prior and the complex speech data distribution.
While the forward process deterministically degrades speech data into Gaussian noise via continuous diffusion, the reverse denoising process lacks a closed-form solution and thus requires a neural network to approximate it.

More recently, flow matching paradigms~\citep{lipman2024fm}, built upon continuous normalizing flows (CNFs)~\citep{torchdiffeq}, have become prevalent in diffusion-based TTS~\citep{le2024voicebox, mehta2024matcha, eskimez2024e2tts, chen2024f5tts}.
CNFs model the transformation as an ordinary differential equation (ODE) and can be efficiently trained using a simulation-free objective known as conditional flow matching (CFM)~\citep{lipman2024fm}.
Although recent studies have demonstrated that DPMs and CFM intrinsically belong to the same theoretical family~\citep{albergo2025stochastic}, CFM is often the preferred choice in practice.
This is because it offers a simpler mathematical formulation~\citep{liu2022recflow}—eliminating the need for complex noise scheduling—while delivering performance comparable or superior to traditional DPMs.

A parallel trajectory in the development of diffusion-based TTS focuses on text-to-speech alignment.
While early systems addressed this challenge by incorporating explicit, auxiliary duration prediction modules~\citep{popov2021gradtts, shen2023ns2, le2024voicebox, ju2024ns3}, recent advances have shifted towards fully end-to-end architectures. For instance, the representative E2-TTS~\citep{eskimez2024e2} framework, along with subsequent studies~\citep{chen2024f5tts, lee2024ditto, zhu2025zipvoice}, demonstrated that the necessary alignment can be implicitly learned by the generative model without explicit supervision, provided there is sufficient training data.

\ttsname{} builds upon this modern trajectory by adopting both the CFM framework and an alignment-free architecture.
However, we extend beyond these foundations by introducing several novel techniques designed to substantially improve the generation quality of diffusion-based TTS.

\subsection{Latent Representations in Diffusion-based TTS}
The choice of latent representation, which serves as the modeling target for the diffusion backbone, is critical in TTS systems.
While it is feasible to train diffusion models directly on raw time-domain waveforms~\citep{gao2023e3tts}, compressing the high-dimensional audio into a compact latent space has proven to be significantly more effective and computationally efficient~\citep{rombach2022ldm}.
Specifically, the latent representation profoundly impacts both generation quality and synthesis speed, as it dictates the inherent trade-off between temporal compression rate and reconstruction fidelity.
Most prior studies have adopted the mel-spectrogram as the default latent representation~\citep{popov2021gradtts, le2024voicebox, eskimez2024e2tts, chen2024f5tts}, necessitating an auxiliary vocoder to invert the predicted mel-spectrograms back into audible waveforms.
To achieve a higher compression rate and further accelerate inference, architectures like DiTTo-TTS~\citep{lee2024ditto} employ a Mel-VAE to encode the mel-spectrograms into an even lower-dimensional space.
However, all these paradigms intrinsically suffer from potential compounding errors.
These errors arise from the multiple stages of data conversion—first predicting the intermediate acoustic features, and subsequently reconstructing the signal via a separate neural vocoder.

In \ttsname{}, we directly employ a waveform-based VAE (Wav-VAE) to encode raw audio into continuous latent representations.
By unifying the acoustic modeling and waveform generation into a single continuous latent space, our approach elegantly bypasses intermediate transformations and mitigates the compounding error problem.

\section{Wav-VAE}
Compared to mel-spectrograms—which inherently discard phase information and fine-grained high-frequency details—compact variational autoencoder (VAE) representations retain essential acoustic characteristics while effectively eliminating redundant components. Consequently, they offer significantly greater potential for high-fidelity audio generation~\citep{liu2022delightfultts, lee2025wave, qiang2024high, niu2025semantic}.

Motivated by these advantages, we develop a fully convolutional audio autoencoder that compresses raw waveforms into a compact, continuous latent representation.
Operating directly in the time domain, the model consists of an encoder $\mathcal{E}$, a bottleneck module, and a decoder $\mathcal{D}$. Given an input waveform $x \in \mathbb{R}^{1 \times T}$, the encoder maps it to a latent sequence $z \in \mathbb{R}^{D \times (T/R)}$, where $D$ denotes the latent dimensionality and $R$ represents the temporal downsampling factor.
Subsequently, the decoder reconstructs the waveform as $\hat{x} = \mathcal{D}(z) \in \mathbb{R}^{1 \times T}$.

\subsection{Model Architecture}
\textbf{Encoder.} The encoder maps the input waveform to a low-dimensional latent sequence via hierarchical downsampling.
The raw waveform is first projected into a high-dimensional feature space using a weight-normalized 1D convolution. The resulting representation is then processed by $N$ cascaded Oobleck blocks~\cite{evans2024fast}. The $i$-th block reduces the temporal resolution by a stride of $s_i$ while expanding the channel dimension from $C_i$ to $C_{i+1}$. The cumulative downsampling ratio is given by:
$
R = \prod_{i=1}^{N} s_i .
$

Prior to downsampling, each block employs a stack of dilated residual units to capture multi-scale temporal dependencies.
A residual unit updates the hidden representation $h$ as follows:
\begin{equation}
h \leftarrow h + 
\mathrm{Conv}_{1\times1}\!\big(
\sigma(\mathrm{Conv}_{k,d}(\sigma(h)))
\big),
\end{equation}
where $\mathrm{Conv}_{k,d}$ denotes a weight-normalized 1D convolution with kernel size $k$ and dilation rate $d$, and $\sigma$ represents the Snake activation function~\citep{ziyin2020neural}.

Following \citet{wu2025clear}, to stabilize the training process under aggressive downsampling, each encoder block incorporates a non-parametric shortcut path. Specifically, let the input to the $i$-th block be a tensor of shape $[B, C_i, T]$ with a target stride of $s_i$.
A space-to-channel reshape operation first folds the temporal dimension into the channel axis, transforming the tensor to $[B, C_i \cdot s_i, T / s_i]$, thereby matching the desired downsampled temporal resolution.
Next, a channel-wise averaging operation groups adjacent channels to reduce the dimension to $C_{i+1}$, yielding a tensor of shape $[B, C_{i+1}, T / s_i]$.
This parameter-free branch establishes a linear residual pathway that bypasses the nonlinear transformations of the main block, and its output is combined with the block's main output via element-wise addition. 

Finally, a convolutional projection layer—also equipped with an analogous shortcut mechanism—is applied to map the deepest features to the target latent dimension $D$.
A VAE bottleneck is then applied to the encoder's output, generating the mean $\mu$ and log-variance $\log \sigma^2$.
The continuous latent representation is sampled using the reparameterization trick: $z = \mu + \sigma \odot \epsilon$, where $\epsilon \sim \mathcal{N}(\mathbf{0}, I)$.

\textbf{Decoder.} The decoder architecture closely mirrors that of the encoder in reverse. The sampled latent sequence $z$ is initially projected into a high-dimensional feature space via a weight-normalized 1D convolution, and then progressively upsampled through $N$ cascaded decoder blocks.
Following each upsampling step, the same stack of dilated residual units used in the encoder is applied to model multi-scale temporal dependencies. 

Furthermore, each decoder block incorporates a non-parametric shortcut branch symmetric to its encoder counterpart.
For an input tensor of shape $[B, C_{i+1}, T / s_i]$, a channel-to-space rearrangement first restores the temporal resolution to $T$.
This is followed by a channel replication step to match the main branch's output shape of $[B, C_i, T]$.
The shortcut and main branch outputs are then fused via element-wise addition.
A final convolutional projection layer maps the reconstructed features back to the time-domain waveform $\hat{x}$.

\subsection{Training Objective}

The Wav-VAE is optimized via a two-stage adversarial training procedure.
The generator (i.e., the autoencoder) minimizes a combined loss function formulated as:
\begin{equation}
    \mathcal{L}_{\mathrm{gen}} = \lambda_{\mathrm{spec}}\mathcal{L}_{\mathrm{spec}} + \lambda_{\mathrm{mel}}\mathcal{L}_{\mathrm{mel}} + \lambda_{\mathrm{time}}\mathcal{L}_{\mathrm{time}} + \lambda_{\mathrm{KL}}\mathcal{L}_{\mathrm{KL}} + \lambda_{\mathrm{adv}}\mathcal{L}_{\mathrm{adv}} + \lambda_{\mathrm{fm}}\mathcal{L}_{\mathrm{fm}}.
\end{equation}
The individual components of this objective are defined as follows:
\begin{itemize}[leftmargin=*, nosep]
    \item $\mathcal{L}_{\mathrm{spec}}$ (Multi-resolution STFT loss~\citep{zeghidour2021soundstream}): Incorporates perceptual weighting to encourage faithful reproduction of the time-frequency structure across various scales.
    \item $\mathcal{L}_{\mathrm{mel}}$ (Multi-scale mel-spectrogram loss~\citep{kumar2023high}): Reduces spectral discrepancies across multiple FFT resolutions, ensuring perceptually natural synthesis.
    \item $\mathcal{L}_{\mathrm{time}}$ (L1 time-domain loss): Directly minimizes the sample-level absolute error between the input and the reconstructed waveforms.
    \item $\mathcal{L}_{\mathrm{KL}}$ (KL divergence loss): Regularizes the learned latent distribution towards a standard Gaussian prior, ensuring a smooth, continuous, and well-structured latent space suitable for the diffusion model.
\end{itemize}

The remaining two terms are derived from a multi-scale STFT discriminator, which is trained in parallel using a standard adversarial objective.
Specifically, the adversarial loss \textbf{$\mathcal{L}_{\mathrm{adv}}$} encourages the generator to synthesize waveforms that are perceptually indistinguishable from real audio.
Meanwhile, the feature matching loss~\citep{Jungil2020hifigan} \textbf{$\mathcal{L}_{\mathrm{fm}}$} minimizes the L1 distance between the intermediate feature maps extracted by the discriminator for both real and reconstructed audio. 

To ensure training stability, we employ an initial warmup phase.
During this period, the adversarial and feature matching terms ($\mathcal{L}_{\mathrm{adv}}$ and $\mathcal{L}_{\mathrm{fm}}$) are disabled.
This strategy allows the autoencoder to establish a stable and accurate reconstruction mapping before being subjected to the more challenging adversarial gradients.
\begin{figure}[t]
    \centering
    \setlength{\belowcaptionskip}{-.5cm}
    \includegraphics[width=0.9\columnwidth]{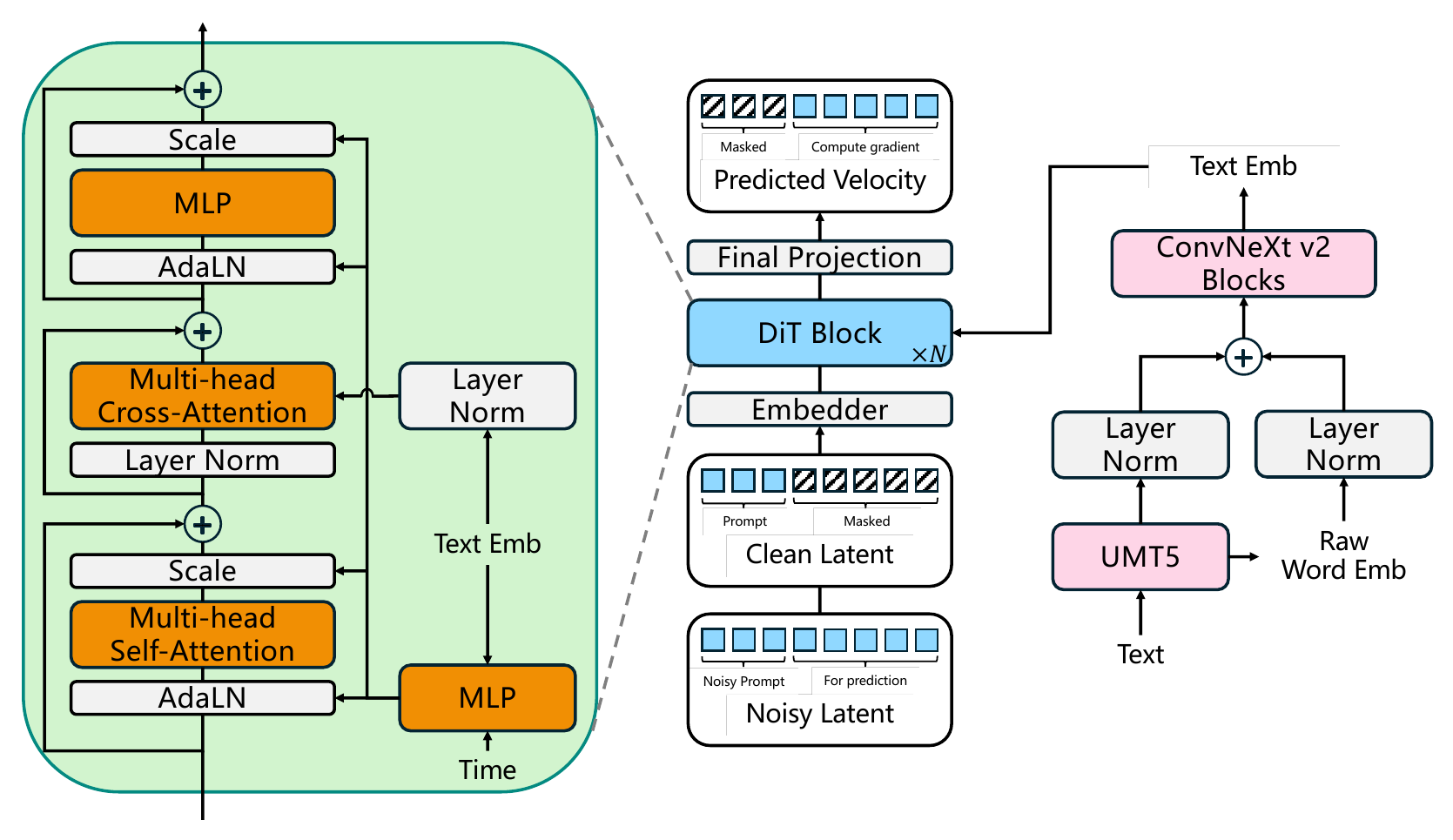}
    \caption{Architecture of \ttsname{}. \emph{Middle}: The overall architecture. \emph{Left}: Detailed structure of the DiT block. \emph{Right}: Detailed structure of the text encoder.}
    \label{figure:architecture}
\end{figure}

\section{Diffusion TTS}
\subsection{Overview}
We adopt the Conditional Flow Matching (CFM) framework~\citep{lipman2024fm} to model the TTS process as an Ordinary Differential Equation (ODE): $dz_{t} = v_{t}dt$, which deterministically transports random Gaussian noise $z_{0}$ to target speech latents $z_{1}$ along a velocity field $v_{t}$.
Following the rectified flow formulation~\citep{liu2022recflow}, we construct the noisy latent $z_{t}$ via linear interpolation between the clean latent and the noise prior:
\begin{equation}
\label{eq:zt}
    z_{t} = (1-t)z_{0} + tz_{1}.
\end{equation}
The velocity field is estimated by a neural network parameterized by $\theta_{\text{CFM}}$, conditioned on the text sequence $q$ and an audio context prompt $z_{ctx}$.
Following VoiceBox~\citep{le2024voicebox}, we construct $z_{ctx}$ by randomly masking continuous spans of the clean latent $z_{1}$, a strategy that inherently enables zero-shot voice cloning capabilities.
The optimization objective for CFM is to minimize the mean squared error between the predicted velocity $v_{\theta}$ and the ground-truth target velocity $(z_{1} - z_{0})$ over the masked regions:
\begin{equation}
\label{eq:cfm_loss}
    \mathcal{L}_{\text{CFM}} = \mathbb{E}_{t, m, z_{0}, z_{1}}\left[ \big\| (1-m) \odot \big((z_{1} - z_{0}) - v(z_{t}, t, z_{ctx}, q; \theta_{\text{CFM}})\big) \big\|^{2} \right],
\end{equation}
where $m$ denotes the random binary mask used to generate $z_{ctx}$.
Furthermore, to facilitate classifier-free guidance (CFG)~\citep{ho2021cfg} during inference, we jointly drop the audio context $z_{ctx}$ and the text condition $q$ with a probability of $10\%$ during training, thereby enabling the model to learn an unconditional distribution.

The overall architecture of our CFM network, illustrated in Fig.~\ref{figure:architecture}, is built upon the Diffusion Transformer (DiT) paradigm~\citep{peebles2023dit}.
It leverages a standard Transformer~\citep{vaswani2017attention} backbone and employs Adaptive Layer Normalization (AdaLN)~\citep{perez2018film} to inject the timestep condition $t$.
To stabilize the training dynamics, we incorporate QK-Norm~\citep{henry2020qknorm} within the attention modules. While standard LayerNorm~\citep{ba2016layernorm} is utilized throughout the network, RMSNorm~\citep{zhang2019rmsnorm} is specifically applied for the QK-Norm operations.
Following DiTTo-TTS~\citep{lee2024ditto}, we utilize cross-attention mechanisms to implicitly learn the text-to-speech alignment, and apply Rotary Positional Embedding (RoPE)~\citep{su2024rope} across all attention layers to capture relative positional dependencies.

We also integrate two structural optimizations from DiTTo-TTS: long-skip connections and a global AdaLN formulation.
The long-skip connection directly adds the network's input to the final-layer hidden state, a modification that yielded slight but consistent improvements in our preliminary experiments.
The global AdaLN mechanism, originally proposed in Gentron~\citep{chen2024gentron}, replaces individual AdaLN projections with a shared, global block for all DiT layers. We observe that this design significantly reduces the overall parameter count without degrading generation performance.

Additionally, we adopt Representation Alignment (REPA)~\citep{yu2024repa} to ground the internal representations of the DiT to a robust, self-supervised semantic space.
Specifically, we employ a pretrained mHuBERT model~\citep{boito2024mhubert} and minimize the L1 distance between the outputs of the $8$-th DiT layer and the corresponding mHuBERT features for the identical input speech.
Our preliminary findings indicate that while REPA does not enhance the generation quality, it substantially accelerates the convergence during training.

In the next section, we detail our text encoder that supports multiple languages.

\subsection{Multilingual Text Embedding}
Our goal is to design a robust text encoder capable of supporting multilingual synthesis.
Existing approaches typically either train a text encoder from scratch~\citep{chen2024f5tts} or leverage a pretrained language model, such as ByT5~\citep{xue2022byt5, lee2024ditto}.
However, training from scratch is highly resource-intensive and notoriously difficult to scale to new languages.
Conversely, while ByT5 theoretically supports arbitrary languages, its byte-level tokenization results in prohibitively long sequence lengths for languages like Chinese, which empirically led to suboptimal performance and alignment difficulties in our preliminary experiments.
To overcome these limitations, we propose utilizing UMT5~\citep{chung2023unimax}, a multilingual variant of T5, as our foundational text encoder.
UMT5 supports $107$ languages and employs a subword tokenizer that maintains reasonable sequence lengths across diverse languages, perfectly aligning with our architectural requirements.
A standard practice when utilizing pretrained language models is to extract the last hidden state as the text representation $q$.
However, we observed that relying exclusively on the final layer yields poor intelligibility in the TTS task.
We hypothesize that while the last hidden state is rich in high-level semantic information, it abstracts away the low-level lexical and phonetic cues that are crucial for precise acoustic mapping.
Motivated by this, we propose integrating the raw word embeddings (the initial embedding layer of UMT5) with the final hidden state.
The resulting text representation $q$ for \ttsname{} is formulated as:
\begin{equation}
    q = \text{LayerNorm}(\text{last\_hidden\_state}) + \text{LayerNorm}(\text{raw\_word\_embedding}).
\end{equation}
Here, non-parametric LayerNorm is applied to appropriately balance the distinct scales of the two representational spaces before summation.
Although our empirical validation is conducted using UMT5, we posit that this dual-embedding extraction strategy is model-agnostic and can be generalized to other large multilingual language models.
We use \texttt{UMT5-base}\footnote{\url{https://huggingface.co/google/umt5-base}} in all experiments.

Furthermore, following F5-TTS~\citep{chen2024f5tts}, we pass the extracted text representation $q$ through a lightweight sequence refinement module based on ConvNeXt V2~\citep{woo2023convnext2}.
We empirically find that this localized convolutional refinement significantly accelerates the convergence of the text-to-speech alignment during training.

In the subsequent sections, we introduce two improvements to the inference process proposed in \ttsname{} that further elevate generation performance.

\subsection{Mitigating the Training-Inference Mismatch in Noisy Latent}
\label{subsection:training_inference_mismatch}
During inference, we employ the Euler method to solve the ODE.
The number of function evaluations is set to $16$.
Initializing the process with randomly sampled Gaussian noise $z_{0}$, we iteratively update the latent $z_{t}$ at each step as follows:
\begin{equation}
    z_{t+\Delta t} = z_{t} + v(z_{t}, t, z_{ctx}, q; \theta_{\text{CFM}}) \Delta t,
\end{equation}
where $\Delta t$ is the predefined integration step size.

By revisiting this sequential inference process, we identify a critical training-inference mismatch regarding the state of the noisy latent $z_{t}$.
For clarity, we conceptually partition $z_{t}$ along the temporal axis into two segments: $z_{t}^{ctx} = z_{t}[:T_{ctx}]$ corresponding to the conditioning prompt, and $z_{t}^{gen} = z_{t}[T_{ctx}:]$ corresponding to the target generation region, where $T_{ctx}$ denotes the duration of the prompt latent $z_{ctx}$.

Recall that during training, the exact trajectory of the entire $z_{t}$ is constructed via linear interpolation (Eq.~\ref{eq:zt}), acting as the ground truth (GT) noisy latent.
During inference, however, an asymmetry emerges.
Because the flow matching objective (Eq.~\ref{eq:cfm_loss}) penalizes velocity prediction errors only on the masked target region ($v^{gen}$), the iterative update successfully yields a valid approximation of the GT trajectory for $z_{t}^{gen}$.
Conversely, because no loss is computed over the prompt region, the model's velocity predictions for $z_{t}^{ctx}$ are essentially unconstrained and arbitrary.
Consequently, accumulating these unconstrained updates causes $z_{t}^{ctx}$ to drift away from its theoretical GT trajectory, thus introducing a training-inference mismatch that has been overlooked in prior work~\citep{le2024voicebox, chen2024f5tts}.
We resolve this discrepancy by forcibly overwriting $z_{t}^{ctx}$ with its GT value at every inference step:
\begin{equation}
    z_{t}^{ctx} \leftarrow t z^{ctx} + (1-t) z_{0}^{ctx},
\end{equation}
where $z_{0}^{ctx}$ is the initial Gaussian noise of the prompt part.

Furthermore, on the basis of this problem, we propose a corollary for CFG.
To obtain a truly unconditional velocity estimate, it is insufficient to merely drop $z_{ctx}$; the explicitly constructed noisy prompt latent $z_{t}^{ctx}$ must also be dropped, as it inherently leaks acoustic information about the prompt.

In Section~\ref{subsubsection:ablation_techniques}, we empirically demonstrate that mitigating this mismatch and isolating the conditional information yields substantial improvements in overall synthesis performance.


\subsection{Replacing CFG with Adaptive Projection Guidance}

Following standard practice, we first utilize classifier-free guidance (CFG)~\citep{ho2021cfg} to steer the predicted velocity at each integration step:
\begin{equation}
\label{eq:cfg}
    v_{t}^{\text{CFG}} = v_{t} + \alpha (v_{t} - v_{t}^{u}),
\end{equation}
where $v_{t}^{u} = v(z_{t}^{u}, t, \varnothing, \varnothing; \theta_{\text{CFM}})$ represents the unconditional velocity; $\alpha$ denotes the CFG scale.
By default, we set $\alpha = 4.0$.
As established in Section~\ref{subsection:training_inference_mismatch}, to accurately compute the unconditional velocity, we compute the noisy latent $z_{t}^{u}$ by dropping the prompt part $z_{t}^{ctx}$ to avoid information leakage, i.e., $z_{t}^{u} = \text{concat}(\varnothing, z_{t}^{gen})$.

In our preliminary experiments, while standard CFG effectively improved synthesis quality, it occasionally introduced audible artifacts, and increasing the guidance scale $\alpha$ further exacerbated the degradation.
We hypothesize that a large CFG scale induces an \textit{oversaturation} phenomenon, a widely recognized issue in diffusion-based image generation~\citep{kynkaanniemi2024cfgig}.
To alleviate this problem, we incorporate Adaptive Projection Guidance (APG)~\citep{sadat2024apg}.
The core intuition of APG is to decompose the guidance residual, $v_{t} - v_{t}^{u}$, into two geometrically orthogonal components: one parallel to the conditional prediction $v_{t}$ and the other orthogonal to it.
APG theorizes that the parallel component is the primary cause behind oversaturation; thus, the issue can be resolved by selectively dampening this term.

To integrate APG into our flow matching framework, we first project the model's output from the velocity domain into the data sample domain (i.e., predicting $z_1$), as suggested by~\citet{sadat2024apg}:
$
    \mu_{t} = z_{t} + (1-t)v_{t}.
$
Let the guidance term in this sample domain be denoted as $\Delta\mu_{t} = \mu_{t} - \mu_{t}^{u}$.
The parallel component $\Delta\mu_{t}^{\parallel}$ with respect to $\mu_{t}$ is calculated as:
$
    \Delta\mu_{t}^{\parallel} = \frac{\langle \Delta\mu_{t}, \mu_{t} \rangle}{\langle \mu_{t}, \mu_{t} \rangle} \mu_{t},
$
and the corresponding orthogonal term is $\Delta\mu_{t}^{\perp} = \Delta\mu_{t} - \Delta\mu_{t}^{\parallel}$.
The APG-adjusted prediction in the sample domain is then formulated as:
\begin{equation}
    \mu_{t}^{\text{APG}} = \mu_{t} + \alpha \Delta\mu_{t}^{\perp} + \eta \Delta\mu_{t}^{\parallel},
\end{equation}
where $\eta$ acts as a dampening factor for the parallel component and is set to $0.5$ by default.
Subsequently, we map the adjusted sample prediction back to the velocity domain to proceed with the ODE solver: 
\begin{equation}
    v_{t}^{\text{APG}} = \frac{\mu_{t}^{\text{APG}} - z_{t}}{1-t}.
\end{equation}
Furthermore, we adopt the reverse momentum trick proposed in APG~\citep{sadat2024apg}, which maintains a moving average $\overline{\Delta\mu_{t}} \leftarrow \Delta\mu_{t} + \beta\overline{\Delta\mu_{t}}$. Applying a negative momentum ($\beta < 0$) forces the guidance to focus more on the current update direction rather than accumulating past momentum.
By default, we set $\beta = -0.3$.

As demonstrated in Section~\ref{subsubsection:ablation_techniques}, APG effectively eliminates artifacts and significantly elevates synthesis quality.
\section{Experiments}
\subsection{Experimental Setup}

\paragraph{Data}
For the training of the Wav-VAE, we employ a curated internal corpus comprising $200$K hours of Chinese and English speech.
Audio clips are segmented to approximately $3$ seconds.

For the TTS backbone (DiT), we utilize a curated internal dataset containing $100$K hours of Chinese and English speech for all baseline and ablation experiments.
For the large-scale scaling experiments, this training corpus is further expanded to $1$M hours.
The transcriptions for all utterances are obtained by a speech recognition model.
We sample all audio data at 24~kHz.
The maximal audio duration-TTS training is 60 seconds.

\paragraph{Training Details}
The Wav-VAE contains $157$M parameters and is optimized on $32$ NVIDIA H800 GPUs with a global batch size of $384$.
By default, the model is configured with a latent dimensionality of $64$ and operates at a temporal frame rate of $11.72$~Hz.

For the diffusion backbone, we train two variants with $1$B and $3.5$B parameters, respectively.
The $1$B model is trained on $16$ GPUs with a global batch size of $256$, whereas the $3.5$B model utilizes $64$ GPUs with a global batch size of $1024$.
Both models are optimized using AdamW~\citep{loshchilov2018adamw}, with moving average coefficients set to $\beta_{1}=0.9$ and $\beta_{2}=0.95$.
We apply a linear learning rate decay schedule, gradually decreasing the learning rate from $1e\mhyphen4$ to $1e\mhyphen5$ following an initial $1$K warmup steps.

\paragraph{Evaluation Metrics}
We benchmark the Wav-VAE on the LibriTTS \texttt{test-clean} subset~\citep{zen2019libritts}, and evaluate the full TTS pipeline on the Seed benchmark~\citep{anastassiou2024seed}.

To evaluate the Wav-VAE reconstruction fidelity, we adopt standard objective metrics including PESQ~\citep{rix2001pesq} for assessing perceptual quality and STOI~\citep{taal2011stoi} for measuring speech intelligibility. 

The generative capabilities of the TTS models are evaluated across four primary dimensions: intelligibility, zero-shot voice cloning, naturalness, and overall acoustic quality.
We measure these using the following metrics:
\begin{itemize}[leftmargin=*, nosep]
    \item \textbf{Character/Word Error Rate (CER/WER):} To quantify intelligibility, we transcribe the synthesized speech using Whisper large-v3~\citep{radford2023whisper} for English and Paraformer~\citep{gao22023funasr} for Chinese, subsequently calculating the respective CER or WER.
    \item \textbf{Speaker Similarity (SIM):} To evaluate voice cloning accuracy, we compute the cosine similarity between the speaker embeddings of the reference prompt and the synthesized speech. This formulation is mathematically equivalent to the SIM-O metric proposed in VoiceBox~\citep{le2024voicebox}. Following Seed-TTS~\citep{anastassiou2024seed}, we utilize a fine-tuned WavLM~\citep{chen2022wavlm} (\texttt{wavlm\_large\_finetune}\footnote{\url{https://github.com/microsoft/UniSpeech/tree/main/downstreams/speaker_verification}}) to extract the robust speaker embeddings.
    \item \textbf{UTMOS~\citep{saeki2022utmos}:} A highly correlated neural objective metric used to approximate human Mean Opinion Scores (MOS) regarding speech naturalness.
    \item \textbf{DNSMOS~\citep{reddy2021dnsmos}:} A widely adopted objective metric designed to evaluate the overall perceptual acoustic quality of the synthesized audio.
\end{itemize}
Note that a subset of these TTS metrics is also applied to evaluate the Wav-VAE reconstructions, allowing us to comparatively analyze the inherent gap between representation reconstruction (Wav-VAE) and generation (TTS).

Finally, we benchmark \ttsname{} against strong prior work, encompassing purely NAR diffusion models, AR models, and state-of-the-art hybrid TTS architectures.

\begin{table}[t]
\centering
\captionsetup{type=table} 
\caption{Objective evaluation results of \ttsname{} on the Seed benchmark~\citep{anastassiou2024seed}. The results of other methods are taken from the original paper or, if open-sourced, evaluated by us. \textbf{Bold} indicates the best score. \underline{Underline} indicates the second-best score.}
\label{tab:seed_benchmark}
\footnotesize
\begin{tabular}{@{}lcccccc@{}}
\toprule
\multirow{2}{*}{\textbf{Model}} & \multicolumn{2}{@{}c}{\textbf{ZH}} & \multicolumn{2}{@{}c}{\textbf{EN}} & \multicolumn{2}{@{}c}{\textbf{ZH-Hard}} \\
 & \textbf{CER (\%)} $\downarrow$ & \textbf{SIM} $\uparrow$ & \textbf{WER (\%)} $\downarrow$ & \textbf{SIM} $\uparrow$ & \textbf{CER (\%)} $\downarrow$ & \textbf{SIM} $\uparrow$ \\
\midrule
GT & 1.26 & 0.755 & 2.14 & 0.734 & - & - \\
\midrule
\multicolumn{5}{@{}l}{\textbf{NAR Models}} \\
Seed-DiT~\citep{anastassiou2024seed} & 1.18 & 0.809  & 1.73 & \textbf{0.790}  & -  & - \\
MaskGCT~\citep{wang2024maskgct} & 2.27 & 0.774 & 2.62 & 0.714  & 10.27 & 0.748 \\
E2 TTS~\citep{eskimez2024e2tts} & 1.97 & 0.730 & 2.19 & 0.710  & - & - \\
F5 TTS~\citep{chen2024f5tts} & 1.56 & 0.741 & 1.83 & 0.647 & 8.67 & 0.713 \\
F5R-TTS~\citep{sun2025f5r} & 1.37 & 0.754 & - & -  & 8.79 & 0.718 \\
ZipVoice~\citep{zhu2025zipvoice} & 1.40 & 0.751 & 1.64 & 0.668 & - & - \\
\midrule
\multicolumn{5}{@{}l}{\textbf{AR/Hybrid Models}} \\
Seed-ICL~\citep{anastassiou2024seed} & 1.12 & 0.796 & 2.25 & 0.762 & 7.59 & 0.776\\
SparkTTS~\citep{wang2025spark} & 1.20 & 0.672 & 1.98 & 0.584 & - & - \\
Qwen2.5-Omni~\citep{xu2025qwen25omni} & 1.70 & 0.752 & 2.72 & 0.632 & 7.97 & 0.747 \\
CosyVoice~\citep{du2024cosyvoice} & 3.63 & 0.723 & 4.29 & 0.609  & 11.75 & 0.709\\
CosyVoice2~\citep{du2024cosyvoice2} & 1.45 & 0.748 & 2.57 & 0.652 & 6.83 & 0.724\\
FireRedTTS-1S~\citep{guo2025fireredtts1s} & 1.05 & 0.750 & 2.17 & 0.660 & 7.63 & 0.748 \\
CosyVoice3-1.5B~\citep{du2025cosyvoice3} & 1.12 & 0.781 & 2.21 & 0.720 & \underline{5.83} & 0.758 \\
IndexTTS2~\citep{zhou2025indextts2} & 1.03 & 0.765 & 2.23 & 0.706  & 7.12 & 0.755 \\
DiTAR~\citep{jia2025ditar}  & 1.02 & 0.753 & 1.69 & 0.735 & - & - \\
MiniMax-Speech~\citep{zhang2025minimax} & 0.99 & 0.799 & 1.90 & 0.738 & - & - \\
VoxCPM~\citep{zhou2025voxcpm}  & \underline{0.93} & 0.772  & 1.85 & 0.729  & 8.87  & 0.730 \\
MOSS-TTS~\citep{team2026moss}  & 1.20 & 0.788  & 1.85 & 0.734  & - & - \\
Qwen3-TTS~\citep{hu2026qwen3tts} & 1.22 & 0.770  & \textbf{1.23} & 0.717  & 6.76  & 0.748 \\
CosyVoice3.5 & \textbf{0.87} & 0.797  & 1.57 & 0.738  & \textbf{5.71} & 0.786 \\
\midrule
\ttsname{}-1B & 1.18  & \underline{0.812}  & 1.78 & 0.762  & 6.33  & \underline{0.787} \\
\ttsname{}-3.5B & 1.09 & \textbf{0.818}  & \underline{1.50} & \underline{0.786} & 6.04  & \textbf{0.797} \\
\bottomrule
\end{tabular}
\end{table}
\subsection{Main Results}
\begin{table}[t]
\caption{Objective evaluation results of the proposed Wav-VAE on the LibriTTS \cite{zen2019libritts} test-clean subset. \textbf{Bold} indicates the best score among continuous VAEs. $N_{q}$ is the number of codebooks for discrete codecs. For codecs, frame per second (FPS) denotes the number of tokens per second.}
\label{tab:wavvae_results}
\centering
\begin{tabular}{@{}lccccc@{}}
\toprule
\textbf{Model} & $\mathbf{N_q}$ & \textbf{FPS} & \textbf{PESQ} $\uparrow$ & \textbf{STOI} $\uparrow$ & \textbf{UTMOS} $\uparrow$ \\
\midrule
GT & -- & -- & 4.644 & 1.0 & 4.056 \\
\midrule
\multicolumn{6}{@{}l}{\textbf{Discrete Codecs}} \\
DAC~\citep{kumar2023high} & 9 & 900 & 3.908 & 0.970 & 3.910 \\
Encodec~\citep{defossez2022encodec} & 8 & 600 & 2.720 & 0.939 & 3.040 \\
Vocos~\citep{siuzdak2023vocos} & 8 & 600 & 2.807 & 0.943 & 3.695 \\
WavTokenizer~\citep{ji2024wavtokenizer} & 1 & 75 & 2.373 & 0.914 & 4.049 \\
BigCodec~\citep{xin2024bigcodec} & 1 & 80 & 2.697 & 0.939 & 4.097 \\
\midrule
\multicolumn{6}{@{}l}{\textbf{Continuous VAEs}} \\
VibeVoice~\citep{peng2025vibevoice} & 1 & 7.50 & 3.068 & 0.828 & \textbf{4.181} \\
\midrule
Ours Wav-VAE & 1 & 7.81  & 3.089 & 0.963 & 4.116 \\
Ours Wav-VAE & 1 & 11.72 & \textbf{3.237} & \textbf{0.967} & 4.013 \\
\bottomrule
\end{tabular}%
\end{table}
The evaluation results for both the full \ttsname{} pipeline and the standalone Wav-VAE are presented in Table~\ref{tab:seed_benchmark} and Table~\ref{tab:wavvae_results}, respectively.

\paragraph{TTS Synthesis Performance} 
As demonstrated in Table~\ref{tab:seed_benchmark}, our proposed TTS model consistently outperforms the majority of prior art, achieving particularly remarkable gains in speaker similarity (SIM) over the highly competitive Seed-DiT architecture~\citep{anastassiou2024seed}.
Specifically, \ttsname{} establishes new state-of-the-art (SOTA) SIM scores on the demanding Seed-ZH and Seed-Hard benchmarks, while securing the second-best SIM score on Seed-EN. 
Most notably, our end-to-end framework decisively surpasses all previous diffusion-based paradigms—such as F5-TTS~\citep{chen2024f5tts}—that rely on intermediate mel-spectrograms as generation targets.
This substantial margin strongly validates our core hypothesis: operating directly within the waveform latent space effectively circumvents compounding errors and yields superior voice cloning fidelity.

Regarding intelligibility (WER/CER), \ttsname{} achieves highly competitive performance relative to existing open-source baselines. While our error rates slightly trail heavily engineered proprietary systems like Qwen3-TTS~\citep{hu2026qwen3tts} and CosyVoice3.5, it is crucial to emphasize that those models rely on complex multi-stage training pipelines and massive amounts of high-quality, human-annotated data.
In contrast, \ttsname{} attains its performance with a remarkably simplified end-to-end architecture and a single training stage.

\paragraph{Wav-VAE Reconstruction Quality}
The intrinsic reconstruction capabilities of our Wav-VAE are detailed in Table~\ref{tab:wavvae_results}.
Operating at a comparable frame rate (FPS), our Wav-VAE exhibits superior overall reconstruction fidelity compared to the baseline Wav-VAE introduced in VibeVoice~\citep{peng2025vibevoice}.
Furthermore, when juxtaposed with SOTA discrete audio codecs, our continuous Wav-VAE not only outperforms most of them in acoustic quality but does so while operating at a drastically reduced sequence length (fewer frames per second).
This stark contrast strongly underscores the inherent capacity advantages and expressive efficiency of modeling continuous latent representations over discrete tokens.

\subsection{Ablation Studies}
To systematically validate our architectural choices and the proposed techniques, we conduct comprehensive ablation experiments. Specifically, our investigations are guided by the following three core research questions (RQs):
\begin{itemize}[leftmargin=*, nosep]
    \item \textbf{RQ1:} As a modeling target-TTS, does the waveform latent (Wav-VAE) outperform intermediate representations like the mel-spectrogram latent (Mel-VAE)?
    \item \textbf{RQ2:} What is the intrinsic relationship between VAE reconstruction fidelity and the downstream TTS synthesis quality? Does a superior VAE guarantee a better generative TTS model?
    \item \textbf{RQ3:} How effectively do our inference techniques, i.e., solving training-inference mismatch and APG, contribute to the overall generation quality?
\end{itemize}

\begin{figure}[t]
    \centering
    \setlength{\belowcaptionskip}{-.5cm}
    \includegraphics[width=0.9\columnwidth]{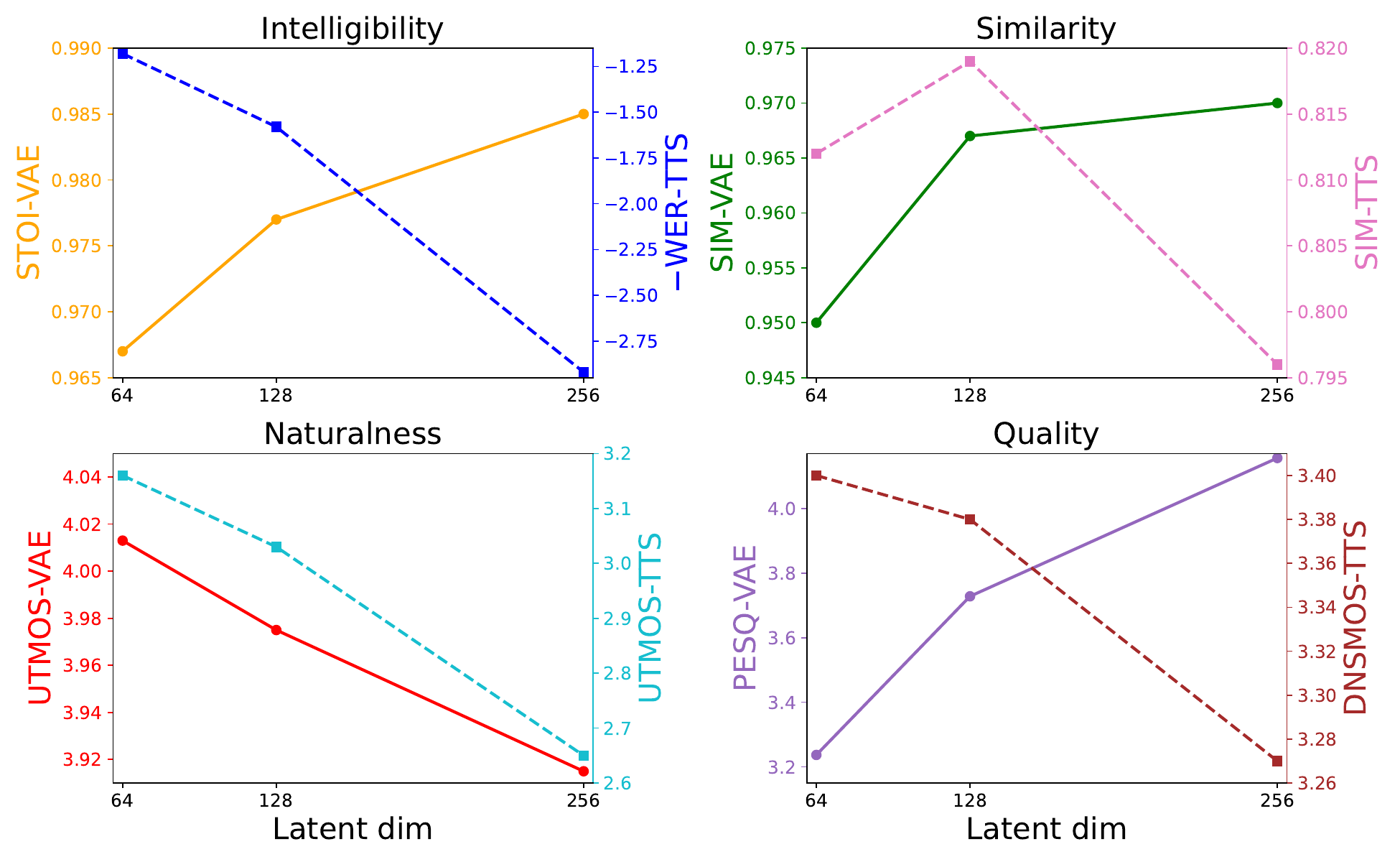}
    \caption{Objective evaluation results for both Wav-VAE reconstruction and TTS synthesis under varying \emph{latent dimensions}. For ease of reading, we negate WER-TTS.}
    \label{figure:ablation_latent_dim_wer_and_sim}
\end{figure}

\subsubsection{RQ1: Wav-VAE vs. Mel-VAE-TTS Generation}

\begin{table}[t]
\centering
\captionsetup{type=table} 
\caption{Objective evaluation results of TTS models based on Wav-VAE and Mel-VAE on the Seed benchmark~\citep{anastassiou2024seed}. \textbf{Bold} indicates the best score.}
\label{tab:tts_ablation_melvae}
\begin{tabular}{@{}ccccccc@{}}
\toprule
\multirow{2}{*}{\textbf{TTS Latent Model}} & \multicolumn{2}{@{}c}{\textbf{ZH}} & \multicolumn{2}{@{}c}{\textbf{EN}} & \multicolumn{2}{@{}c}{\textbf{ZH-Hard}} \\
 & \textbf{CER (\%)} $\downarrow$ & \textbf{SIM} $\uparrow$ & \textbf{WER (\%)} $\downarrow$ & \textbf{SIM} $\uparrow$ & \textbf{CER (\%)} $\downarrow$ & \textbf{SIM} $\uparrow$ \\
\midrule
Mel-VAE & 1.29 & 0.706  & 2.20 & 0.714  & 7.70  & 0.696 \\
Wav-VAE & \textbf{1.18}  & \textbf{0.812}  & \textbf{1.78} & \textbf{0.762}  & \textbf{6.33}  & \textbf{0.787} \\
\bottomrule
\end{tabular}
\end{table}
The central hypothesis underpinning \ttsname{} is that modeling directly within the waveform latent space is superior to utilizing intermediate representations, primarily due to the mitigation of compounding errors.
Since recent work like DiTTo-TTS~\citep{lee2024ditto} has already established that Mel-VAE outperforms raw mel-spectrograms in diffusion-based TTS, we restrict our comparison directly to Wav-VAE versus Mel-VAE.

For this experiment, we adopt the open-source Mel-VAE introduced in ACE-Step~\citep{gong2025acestep}.
Although originally designed for music generation, we empirically verify that this Mel-VAE yields high-fidelity speech reconstruction at a similar frame rate to our proposed Wav-VAE.
We train a baseline $1$B parameter TTS model using this Mel-VAE as the modeling target. During inference, the generated latents are decoded into mel-spectrograms, which are subsequently inverted into time-domain waveforms using the officially provided high-quality vocoder\footnote{\url{https://github.com/ace-step/ACE-Step}}.

The comparative evaluation results are presented in Table~\ref{tab:tts_ablation_melvae}. 
As observed, the \ttsname{} model built upon the Wav-VAE consistently and significantly outperforms the Mel-VAE-based baseline across all metrics, validating our core assumption.
Remarkably, while improvements in intelligibility (WER/CER) are solid, the Wav-VAE yields a drastic boost in the speaker similarity (SIM) metric.
This targeted improvement elegantly corroborates our hypothesis: fine-grained, high-frequency acoustic details—which are essential for zero-shot voice cloning—are intrinsically fragile and easily lost during the cascading conversions (latent $\rightarrow$ mel-spectrogram $\rightarrow$ waveform) inherent to the Mel-VAE pipeline.

\begin{figure}[t]
    \centering
    \setlength{\belowcaptionskip}{-.5cm}
    \includegraphics[width=0.9\columnwidth]{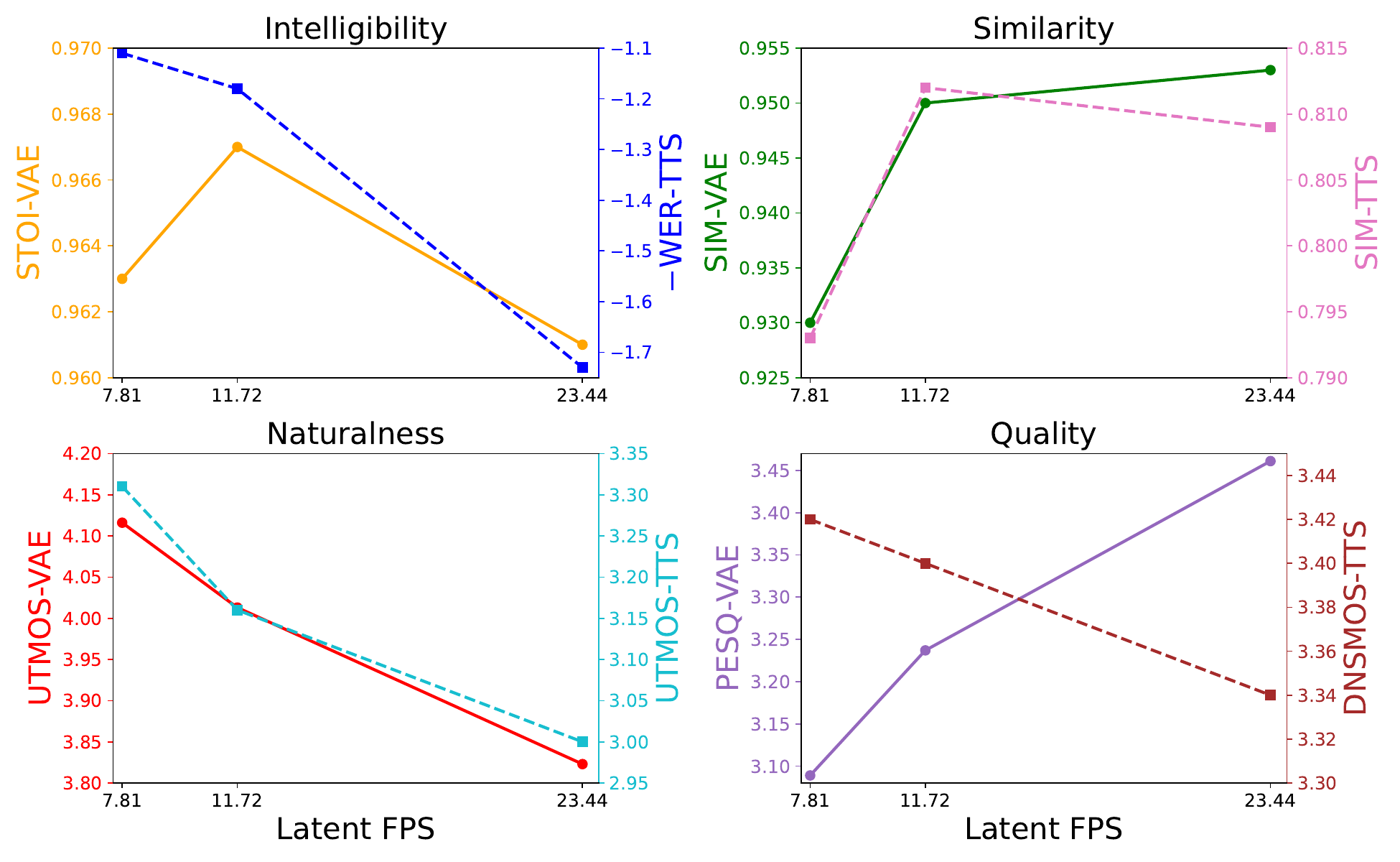}
    \caption{Objective evaluation results for both Wav-VAE reconstruction and TTS synthesis across varying \emph{latent frame rates (FPS)}. For ease of reading, we negate WER-TTS.}
    \label{figure:ablation_latent_fps_wer_and_sim}
\end{figure}

\subsubsection{RQ2: The Interplay Between Wav-VAE Reconstruction and TTS Generation}

We investigate the intrinsic relationship between the reconstruction fidelity of the Wav-VAE and the generation quality of the downstream TTS model.
A naive assumption is that a superior Wav-VAE guarantees better TTS performance, given that the VAE's reconstruction fidelity inherently defines the upper bound for the generative model.
To test this hypothesis, we train multiple Wav-VAEs with varying latent dimensionalities and temporal frame rates (FPS), subsequently training a corresponding TTS backbone for each VAE variant.
Specifically, we select latent dimensions from the set $\{64,\ 128,\ 256\}$ and frame rates from $\{7.81,\ 11.72,\ 23.44\}$, yielding a total of $6$ unique Wav-VAE models and $6$ paired TTS models.
For the dimension ablation (3 models), we fix the frame rate at $20$ Hz; conversely, for the frame rate ablation (3 models), we fix the latent dimension at $64$.
All TTS models in this ablation are trained using the exact configurations as the \ttsname{}-1B baseline.

The comprehensive evaluation results are visualized in Fig.~\ref{figure:ablation_latent_dim_wer_and_sim} and Fig.~\ref{figure:ablation_latent_fps_wer_and_sim}.
To facilitate a clear comparison across domains, we categorize the metrics into four analogous groups: intelligibility (STOI-VAE \& WER-TTS), speaker similarity (SIM-VAE \& SIM-TTS), naturalness (UTMOS-VAE \& UTMOS-TTS), and overall acoustic quality (PESQ-VAE \& DNSMOS-TTS).
Note that the VAE similarity (SIM-VAE) is calculated by comparing the ground truth (GT) utterance against its direct reconstruction.

\textbf{Observation 1: The Dimension-Capacity Trade-off.} 
\emph{Under a fixed TTS parameter budget, increasing the latent dimension consistently improves the Wav-VAE's reconstruction fidelity but simultaneously degrades the TTS generation quality (see Fig.~\ref{figure:ablation_latent_dim_wer_and_sim}).} 
This finding directly contradicts the naive assumption.
We initially hypothesized that increasing the TTS model capacity might resolve this mismatch; thus, we scaled up the TTS backbone to $3.5$B parameters, conditioned on the $128$-dimensional Wav-VAE.
However, while this larger variant achieved a marginal gain in SIM score, its overall performance remained inferior to the $3.5$B model conditioned on the $64$-dimensional Wav-VAE (as reported in Table~\ref{tab:seed_benchmark}).
This suggests that excessively high-dimensional continuous latents impose a severe modeling burden on the diffusion backbone that cannot be easily overcome merely by scaling up parameters.

\textbf{Observation 2: The Frame Rate Sweet Spot.} 
\emph{There exists an optimal temporal frame rate (FPS) that balances VAE and TTS performance, though this sweet spot is not necessarily identical for both tasks (see Fig.~\ref{figure:ablation_latent_fps_wer_and_sim}).}
For the Wav-VAE, a lower FPS surprisingly yields better intelligibility and naturalness, but penalizes similarity and overall acoustic quality.
This behavior is intuitive: an aggressively downsampled (lower FPS) latent forces the autoencoder to discard fine-grained, high-frequency acoustic details (hurting SIM and PESQ) while preserving global phonetic structures (aiding STOI). 
Conversely, for the generative TTS model, a lower FPS substantially boosts the overall synthesis quality.
We observe that the diffusion backbone struggles to accurately model the complex, highly correlated temporal dynamics of high-FPS latents, leading to unstable generation. 

Synthesizing these two critical observations, we empirically identify the $64$-dimensional, $11.72$-Hz Wav-VAE as the optimal representation target, and adopt it as the default configuration for all \ttsname{} models.

\subsubsection{RQ3: Effectiveness of the Proposed Techniques for Inference}
\label{subsubsection:ablation_techniques}
\begin{table}[t]
\centering
\captionsetup{type=table} 
\caption{Objective evaluation results of the ablation studies on noise-prompt dual masking and APG on the Seed-ZH benchmark~\citep{anastassiou2024seed}. \textbf{Bold} indicates the best score.}
\label{tab:tts_ablation_other_techniques}
\begin{tabular}{@{}lcccc@{}}
\toprule
\textbf{Experiment}  & \textbf{CER (\%)} $\downarrow$ & \textbf{SIM} $\uparrow$ & \textbf{UTMOS} $\uparrow$ & \textbf{DNSMOS} $\uparrow$ \\
\midrule
\ttsname{}-1B & \textbf{1.18} & \textbf{0.812} & \textbf{3.16} & \textbf{3.40}\\
\midrule
training-inference mismatch & 1.21 & 0.769  & 2.83 & 3.34 \\
w/o APG & \textbf{1.18} & \textbf{0.812}  & 3.06 & 3.38\\
\bottomrule
\end{tabular}
\end{table}
Finally, we address RQ3 by evaluating the individual contributions of solving the training-inference mismatch and APG.
To this end, we conduct two targeted ablation experiments on the \ttsname{}-1B backbone.
In the first configuration (\textit{training-inference mismatch}), we keep $z_{t}^{ctx}$ as the model prediction and do not overwrite it with the GT noisy latent for inference.
We also retain $z_{t}^{ctx}$ to compute the unconditional velocity.
In the second configuration (\textit{w/o APG}), we replace the APG inference algorithm with standard CFG (Eq.~\ref{eq:cfg}).
The comparative results are summarized in Table~\ref{tab:tts_ablation_other_techniques}.
\begin{itemize}[leftmargin=*, nosep]
    \item \textbf{Impact of the training-inference mismatch:} The overall performance of the utterances synthesized by \ttsname{}-1B consistently and significantly outperforms those synthesized without solving the training-inference mismatch problem. This clear performance degradation validates the existence of the recognized problem and the effectiveness of our method to mitigate it.
    \item \textbf{Impact of APG:} While the baseline model employing standard CFG achieves comparable intelligibility (CER) and speaker similarity (SIM) scores, the integration of APG yields superior UTMOS and DNSMOS scores. This demonstrates that APG effectively mitigates the oversaturation artifacts inherent to high-scale CFG, thereby elevating the perceptual naturalness and overall acoustic quality of the synthesized speech.
\end{itemize}
\section{Conclusion and Future Work}
In this paper, we present \ttsname{}, a state-of-the-art non-autoregressive diffusion-based TTS model. 
The core advancement of \ttsname{} lies in modeling the generative process directly within the waveform latent space, bypassing intermediate acoustic representations such as mel-spectrograms widely adopted in prior literature. 
This unified design not only drastically simplifies the overall TTS pipeline but also fundamentally eliminates the compounding errors inherently caused by two-stage acoustic-to-waveform conversions. 
Furthermore, we introduce two critical improvements to the inference process: first, we identify and rectify a long-standing training-inference mismatch; second, we replace traditional CFG with APG to elevate generation quality.

Extensive experimental results demonstrate that \ttsname{} achieves new SOTA zero-shot speaker similarity on the rigorous Seed benchmark while maintaining competitive intelligibility. Notably, this is accomplished through an end-to-end approach, without relying on sophisticated multi-stage training pipelines or expensive high-quality human annotations. 
By outperforming previous diffusion-based baselines by a considerable margin, our work robustly validates the superiority of waveform-level latent modeling over traditional intermediate representations.

Finally, through comprehensive ablation studies, we systematically dissect the individual contributions of our proposed components. 
Most importantly, our deep dive into the interplay between the Wav-VAE's reconstruction fidelity (e.g., varying dimensions and frame rates) and the downstream TTS generation quality reveals non-trivial trade-offs.
We believe these empirical insights advance the understanding of the synergy between representation learning and generative modeling, shedding light on the future design of audio foundation models.

\paragraph{Future Work} Promising directions for future research include pushing the performance ceiling via alignment-free reinforcement learning (RLHF for audio), and accelerating the inference speed through knowledge distillation techniques for real-time deployment.
\section{Contributor}
\subsection*{Core Contributors}
Detai Xin, Shujie Hu, Chengzuo Yang

\subsection*{Tech Leads}
Chen Huang, Guoqiao Yu, Guanglu Wan, Xunliang Cai

\subsection*{Contributors}
(\emph{Sorted in alphabetical order})\\
Disong Wang, Fengjiao Chen, Fengyu Yang, Hui Yang, Jiamu Li, Jun Wang, Qi Li, Qian Yang, Quanxiu Wang, Rumei Li, Shuaiqi Chen, Xu Xiang, Xuezhi Cao, Yi Chen, Yuchen Sun, Zheng Zhang, Zhiqing Hong, Ziwen Wang

\bibliographystyle{unsrtnat}
\bibliography{ref}

@string{is =  "Proc. Interspeech"}

@string{icassp="Proc. ICASSP"}

@article{vaswani2017attention,
  title={Attention is all you need},
  author={Vaswani, Ashish and Shazeer, Noam and Parmar, Niki and Uszkoreit, Jakob and Jones, Llion and Gomez, Aidan N and Kaiser, {\L}ukasz and Polosukhin, Illia},
  journal={Advances in neural information processing systems},
  volume={30},
  year={2017}
}

@inproceedings{popov2021gradtts,
  title={Grad-tts: A diffusion probabilistic model for text-to-speech},
  author={Popov, Vadim and Vovk, Ivan and Gogoryan, Vladimir and Sadekova, Tasnima and Kudinov, Mikhail},
  booktitle={International Conference on Machine Learning},
  pages={8599--8608},
  year={2021},
  organization={PMLR}
}

@article{shen2023ns2,
  title={Naturalspeech 2: Latent diffusion models are natural and zero-shot speech and singing synthesizers},
  author={Shen, Kai and Ju, Zeqian and Tan, Xu and Liu, Yanqing and Leng, Yichong and He, Lei and Qin, Tao and Zhao, Sheng and Bian, Jiang},
  journal={arXiv preprint arXiv:2304.09116},
  year={2023}
}

@article{wang2023valle,
  title={Neural codec language models are zero-shot text to speech synthesizers},
  author={Wang, Chengyi and Chen, Sanyuan and Wu, Yu and Zhang, Ziqiang and Zhou, Long and Liu, Shujie and Chen, Zhuo and Liu, Yanqing and Wang, Huaming and Li, Jinyu and others},
  journal={arXiv preprint arXiv:2301.02111},
  year={2023}
}

@article{zeghidour2021soundstream,
  title={Soundstream: An end-to-end neural audio codec},
  author={Zeghidour, Neil and Luebs, Alejandro and Omran, Ahmed and Skoglund, Jan and Tagliasacchi, Marco},
  journal={IEEE/ACM Transactions on Audio, Speech, and Language Processing},
  volume={30},
  pages={495--507},
  year={2021},
  publisher={IEEE}
}

@article{defossez2022encodec,
  title={High fidelity neural audio compression},
  author={D{\'e}fossez, Alexandre and Copet, Jade and Synnaeve, Gabriel and Adi, Yossi},
  journal={arXiv preprint arXiv:2210.13438},
  year={2022}
}

@article{peng2025vibevoice,
  title={Vibevoice technical report},
  author={Peng, Zhiliang and Yu, Jianwei and Wang, Wenhui and Chang, Yaoyao and Sun, Yutao and Dong, Li and Zhu, Yi and Xu, Weijiang and Bao, Hangbo and Wang, Zehua and others},
  journal={arXiv preprint arXiv:2508.19205},
  year={2025}
}

@article{ji2024wavtokenizer,
  title={Wavtokenizer: an efficient acoustic discrete codec tokenizer for audio language modeling},
  author={Ji, Shengpeng and Jiang, Ziyue and Wang, Wen and Chen, Yifu and Fang, Minghui and Zuo, Jialong and Yang, Qian and Cheng, Xize and Wang, Zehan and Li, Ruiqi and others},
  journal={arXiv preprint arXiv:2408.16532},
  year={2024}
}

@article{siuzdak2023vocos,
  title={Vocos: Closing the gap between time-domain and fourier-based neural vocoders for high-quality audio synthesis},
  author={Siuzdak, Hubert},
  journal={arXiv preprint arXiv:2306.00814},
  year={2023}
}

@article{kumar2023high,
  title={High-fidelity audio compression with improved rvqgan},
  author={Kumar, Rithesh and Seetharaman, Prem and Luebs, Alejandro and Kumar, Ishaan and Kumar, Kundan},
  journal={Advances in Neural Information Processing Systems},
  volume={36},
  pages={27980--27993},
  year={2023}
}

@article{le2024voicebox,
  title={Voicebox: Text-guided multilingual universal speech generation at scale},
  author={Le, Matthew and Vyas, Apoorv and Shi, Bowen and Karrer, Brian and Sari, Leda and Moritz, Rashel and Williamson, Mary and Manohar, Vimal and Adi, Yossi and Mahadeokar, Jay and others},
  journal={Advances in neural information processing systems},
  volume={36},
  year={2024}
}

@article{saeki2022utmos,
  title={UTMOS: Utokyo-sarulab system for voiceMOS Challenge 2022},
  author={Saeki, Takaaki and Xin, Detai and Nakata, Wataru and Koriyama, Tomoki and Takamichi, Shinnosuke and Saruwatari, Hiroshi},
  journal={arXiv preprint arXiv:2204.02152},
  year={2022}
}

@inproceedings{loshchilov2018adamw,
  title={Decoupled Weight Decay Regularization},
  author={Loshchilov, Ilya and Hutter, Frank},
  booktitle={Proc. ICLR},
  year={2018}
}

@article{chen2022wavlm,
  title={Wavlm: Large-scale self-supervised pre-training for full stack speech processing},
  author={Chen, Sanyuan and Wang, Chengyi and Chen, Zhengyang and Wu, Yu and Liu, Shujie and Chen, Zhuo and Li, Jinyu and Kanda, Naoyuki and Yoshioka, Takuya and Xiao, Xiong and others},
  journal={IEEE Journal of Selected Topics in Signal Processing},
  volume={16},
  number={6},
  pages={1505--1518},
  year={2022},
  publisher={IEEE}
}

@article{ju2024ns3,
  title={NaturalSpeech 3: Zero-Shot Speech Synthesis with Factorized Codec and Diffusion Models},
  author={Ju, Zeqian and Wang, Yuancheng and Shen, Kai and Tan, Xu and Xin, Detai and Yang, Dongchao and Liu, Yanqing and Leng, Yichong and Song, Kaitao and Tang, Siliang and others},
  journal={arXiv preprint arXiv:2403.03100},
  year={2024}
}

@article{zen2019libritts,
  title={LibriTTS: A Corpus Derived from LibriSpeech for Text-to-Speech},
  author={Zen, Heiga and Dang, Viet and Clark, Rob and Zhang, Yu and Weiss, Ron J and Jia, Ye and Chen, Zhifeng and Wu, Yonghui},
  journal={Proc. Interspeech},
  year={2019},
  publisher={ISCA}
}

@article{ren2019fastspeech,
  title={Fastspeech: Fast, robust and controllable text to speech},
  author={Ren, Yi and Ruan, Yangjun and Tan, Xu and Qin, Tao and Zhao, Sheng and Zhao, Zhou and Liu, Tie-Yan},
  journal={Proc. NeurIPS},
  volume={32},
  year={2019}
}

@article{betker2023tortoise,
  title={Better speech synthesis through scaling},
  author={Betker, James},
  journal={arXiv preprint arXiv:2305.07243},
  year={2023}
}

@article{du2024cosyvoice2,
  title={Cosyvoice 2: Scalable streaming speech synthesis with large language models},
  author={Du, Zhihao and Wang, Yuxuan and Chen, Qian and Shi, Xian and Lv, Xiang and Zhao, Tianyu and Gao, Zhifu and Yang, Yexin and Gao, Changfeng and Wang, Hui and others},
  journal={arXiv preprint arXiv:2412.10117},
  year={2024}
}

@article{du2024cosyvoice,
  title={Cosyvoice: A scalable multilingual zero-shot text-to-speech synthesizer based on supervised semantic tokens},
  author={Du, Zhihao and Chen, Qian and Zhang, Shiliang and Hu, Kai and Lu, Heng and Yang, Yexin and Hu, Hangrui and Zheng, Siqi and Gu, Yue and Ma, Ziyang and others},
  journal={arXiv preprint arXiv:2407.05407},
  year={2024}
}

@article{du2025cosyvoice3,
  title={Cosyvoice 3: Towards in-the-wild speech generation via scaling-up and post-training},
  author={Du, Zhihao and Gao, Changfeng and Wang, Yuxuan and Yu, Fan and Zhao, Tianyu and Wang, Hao and Lv, Xiang and Wang, Hui and Ni, Chongjia and Shi, Xian and others},
  journal={arXiv preprint arXiv:2505.17589},
  year={2025}
}

@article{anastassiou2024seed,
  title={Seed-tts: A family of high-quality versatile speech generation models},
  author={Anastassiou, Philip and Chen, Jiawei and Chen, Jitong and Chen, Yuanzhe and Chen, Zhuo and Chen, Ziyi and Cong, Jian and Deng, Lelai and Ding, Chuang and Gao, Lu and others},
  journal={arXiv preprint arXiv:2406.02430},
  year={2024}
}

@article{zhou2025indextts2,
  title={IndexTTS2: A Breakthrough in Emotionally Expressive and Duration-Controlled Auto-Regressive Zero-Shot Text-to-Speech},
  author={Zhou, Siyi and Zhou, Yiquan and He, Yi and Zhou, Xun and Wang, Jinchao and Deng, Wei and Shu, Jingchen},
  journal={arXiv preprint arXiv:2506.21619},
  year={2025}
}

@article{xu2025qwen25omni,
  title={Qwen2. 5-omni technical report},
  author={Xu, Jin and Guo, Zhifang and He, Jinzheng and Hu, Hangrui and He, Ting and Bai, Shuai and Chen, Keqin and Wang, Jialin and Fan, Yang and Dang, Kai and others},
  journal={arXiv preprint arXiv:2503.20215},
  year={2025}
}

@article{su2024rope,
  title={Roformer: Enhanced transformer with rotary position embedding},
  author={Su, Jianlin and Ahmed, Murtadha and Lu, Yu and Pan, Shengfeng and Bo, Wen and Liu, Yunfeng},
  journal={Neurocomputing},
  volume={568},
  pages={127063},
  year={2024},
  publisher={Elsevier}
}

@article{chen2024f5tts,
  title={F5-tts: A fairytaler that fakes fluent and faithful speech with flow matching},
  author={Chen, Yushen and Niu, Zhikang and Ma, Ziyang and Deng, Keqi and Wang, Chunhui and Zhao, Jian and Yu, Kai and Chen, Xie},
  journal={arXiv preprint arXiv:2410.06885},
  year={2024}
}

@inproceedings{perez2018film,
  title={Film: Visual reasoning with a general conditioning layer},
  author={Perez, Ethan and Strub, Florian and De Vries, Harm and Dumoulin, Vincent and Courville, Aaron},
  booktitle={Proceedings of the AAAI conference on artificial intelligence},
  volume={32},
  year={2018}
}

@inproceedings{ho2021cfg,
  title={Classifier-Free Diffusion Guidance},
  author={Ho, Jonathan and Salimans, Tim},
  year={2021},
  booktitle={NeurIPS 2021 Workshop on Deep Generative Models and Downstream Applications}
}

@article{liu2022recflow,
  title={Flow straight and fast: Learning to generate and transfer data with rectified flow},
  author={Liu, Xingchao and Gong, Chengyue and Liu, Qiang},
  journal={arXiv preprint arXiv:2209.03003},
  year={2022}
}

@inproceedings{lipman2024fm,
  title={Flow Matching for Generative Modeling},
  author={Lipman, Yaron and Chen, Ricky TQ and Ben-Hamu, Heli and Nickel, Maximilian and Le, Matthew},
  year={2022},
  booktitle={The Eleventh International Conference on Learning Representations}
}

@article{sun2025f5r,
  title={F5R-TTS: Improving flow-matching based text-to-speech with group relative policy optimization},
  author={Sun, Xiaohui and Xiao, Ruitong and Mo, Jianye and Wu, Bowen and Yu, Qun and Wang, Baoxun},
  journal={arXiv preprint arXiv:2504.02407},
  year={2025}
}

@article{wang2025spark,
  title={Spark-tts: An efficient llm-based text-to-speech model with single-stream decoupled speech tokens},
  author={Wang, Xinsheng and Jiang, Mingqi and Ma, Ziyang and Zhang, Ziyu and Liu, Songxiang and Li, Linqin and Liang, Zheng and Zheng, Qixi and Wang, Rui and Feng, Xiaoqin and others},
  journal={arXiv preprint arXiv:2503.01710},
  year={2025}
}

@article{ho2020ddpm,
  title={Denoising diffusion probabilistic models},
  author={Ho, Jonathan and Jain, Ajay and Abbeel, Pieter},
  journal={Advances in neural information processing systems},
  volume={33},
  pages={6840--6851},
  year={2020}
}

@article{ba2016layernorm,
  title={Layer normalization},
  author={Ba, Jimmy Lei and Kiros, Jamie Ryan and Hinton, Geoffrey E},
  journal={arXiv preprint arXiv:1607.06450},
  year={2016}
}

@article{zhang2019rmsnorm,
  title={Root mean square layer normalization},
  author={Zhang, Biao and Sennrich, Rico},
  journal={Advances in neural information processing systems},
  volume={32},
  year={2019}
}

@inproceedings{reddy2021dnsmos,
  title={DNSMOS: A non-intrusive perceptual objective speech quality metric to evaluate noise suppressors},
  author={Reddy, Chandan KA and Gopal, Vishak and Cutler, Ross},
  booktitle={ICASSP 2021-2021 IEEE International Conference on Acoustics, Speech and Signal Processing (ICASSP)},
  pages={6493--6497},
  year={2021},
  organization={IEEE}
}

@inproceedings{Jungil2020hifigan,
 author = {Kong, Jungil and Kim, Jaehyeon and Bae, Jaekyoung},
 booktitle = {Advances in Neural Information Processing Systems},
 pages = {17022--17033},
 publisher = {Curran Associates, Inc.},
 title = {HiFi-GAN: Generative Adversarial Networks for Efficient and High Fidelity Speech Synthesis},
 volume = {33},
 year = {2020}
}

@inproceedings{gao22023funasr,
  title     = {FunASR: A Fundamental End-to-End Speech Recognition Toolkit},
  author    = {Zhifu Gao and Zerui Li and Jiaming Wang and Haoneng Luo and Xian Shi and Mengzhe Chen and Yabin Li and Lingyun Zuo and Zhihao Du and Shiliang Zhang},
  year      = {2023},
  booktitle = {Interspeech 2023},
  pages     = {1593--1597},
  doi       = {10.21437/Interspeech.2023-1428},
  issn      = {2958-1796},
}

@inproceedings{radford2023whisper,
  title={Robust speech recognition via large-scale weak supervision},
  author={Radford, Alec and Kim, Jong Wook and Xu, Tao and Brockman, Greg and McLeavey, Christine and Sutskever, Ilya},
  booktitle={International conference on machine learning},
  pages={28492--28518},
  year={2023},
  organization={PMLR}
}

@article{wang2024maskgct,
  title={Maskgct: Zero-shot text-to-speech with masked generative codec transformer},
  author={Wang, Yuancheng and Zhan, Haoyue and Liu, Liwei and Zeng, Ruihong and Guo, Haotian and Zheng, Jiachen and Zhang, Qiang and Zhang, Xueyao and Zhang, Shunsi and Wu, Zhizheng},
  journal={arXiv preprint arXiv:2409.00750},
  year={2024}
}

@inproceedings{eskimez2024e2tts,
  title={E2 tts: Embarrassingly easy fully non-autoregressive zero-shot tts},
  author={Eskimez, Sefik Emre and Wang, Xiaofei and Thakker, Manthan and Li, Canrun and Tsai, Chung-Hsien and Xiao, Zhen and Yang, Hemin and Zhu, Zirun and Tang, Min and Tan, Xu and others},
  booktitle={2024 IEEE Spoken Language Technology Workshop (SLT)},
  pages={682--689},
  year={2024},
  organization={IEEE}
}

@article{zhang2025minimax,
  title={Minimax-speech: Intrinsic zero-shot text-to-speech with a learnable speaker encoder},
  author={Zhang, Bowen and Guo, Congchao and Yang, Geng and Yu, Hang and Zhang, Haozhe and Lei, Heidi and Mai, Jialong and Yan, Junjie and Yang, Kaiyue and Yang, Mingqi and others},
  journal={arXiv preprint arXiv:2505.07916},
  year={2025}
}

@article{guo2025fireredtts1s,
  title={Fireredtts-1s: An upgraded streamable foundation text-to-speech system},
  author={Guo, Hao-Han and Hu, Yao and Shen, Fei-Yu and Tang, Xu and Wu, Yi-Chen and Xie, Feng-Long and Xie, Kun},
  journal={arXiv preprint arXiv:2503.20499},
  year={2025}
}

@article{lee2024ditto,
  title={DiTTo-TTS: Diffusion transformers for scalable text-to-speech without domain-specific factors},
  author={Lee, Keon and Kim, Dong Won and Kim, Jaehyeon and Chung, Seungjun and Cho, Jaewoong},
  journal={arXiv preprint arXiv:2406.11427},
  year={2024}
}

@article{zhu2025zipvoice,
  title={Zipvoice: Fast and high-quality zero-shot text-to-speech with flow matching},
  author={Zhu, Han and Kang, Wei and Yao, Zengwei and Guo, Liyong and Kuang, Fangjun and Li, Zhaoqing and Zhuang, Weiji and Lin, Long and Povey, Daniel},
  journal={arXiv preprint arXiv:2506.13053},
  year={2025}
}

@inproceedings{eskimez2024e2,
  title={E2 tts: Embarrassingly easy fully non-autoregressive zero-shot tts},
  author={Eskimez, Sefik Emre and Wang, Xiaofei and Thakker, Manthan and Li, Canrun and Tsai, Chung-Hsien and Xiao, Zhen and Yang, Hemin and Zhu, Zirun and Tang, Min and Tan, Xu and others},
  booktitle={2024 IEEE spoken language technology workshop (SLT)},
  pages={682--689},
  year={2024},
  organization={IEEE}
}

@article{hu2026qwen3tts,
  title={Qwen3-TTS Technical Report},
  author={Hu, Hangrui and Zhu, Xinfa and He, Ting and Guo, Dake and Zhang, Bin and Wang, Xiong and Guo, Zhifang and Jiang, Ziyue and Hao, Hongkun and Guo, Zishan and others},
  journal={arXiv preprint arXiv:2601.15621},
  year={2026}
}

@article{kingma2013vae,
  title={Auto-encoding variational bayes},
  author={Kingma, Diederik P and Welling, Max},
  journal={arXiv preprint arXiv:1312.6114},
  year={2013}
}

@inproceedings{peebles2023dit,
  title={Scalable diffusion models with transformers},
  author={Peebles, William and Xie, Saining},
  booktitle={Proceedings of the IEEE/CVF international conference on computer vision},
  pages={4195--4205},
  year={2023}
}

@inproceedings{liu2022delightfultts,
  title={Delightfultts 2: End-to-end speech synthesis with adversarial vector-quantized auto-encoders},
  author={Liu, Yanqing and Xue, Ruiqing and He, Lei and Tan, Xu and Zhao, Sheng},
  booktitle=is,
  year={2022}
}

@inproceedings{lee2025wave,
  title={Wave-u-mamba: an end-to-end framework for high-quality and efficient speech super resolution},
  author={Lee, Yongjoon and Kim, Chanwoo},
  booktitle=icassp,
  year={2025}
}

@inproceedings{qiang2024high,
  title={High-fidelity speech synthesis with minimal supervision: All using diffusion models},
  author={Qiang, Chunyu and Li, Hao and Tian, Yixin and Zhao, Yi and others},
  booktitle=icassp,
  year={2024}
}

@article{niu2025semantic,
  title={Semantic-VAE: Semantic-Alignment Latent Representation for Better Speech Synthesis},
  author={Niu, Zhikang and Hu, Shujie and Choi, Jeongsoo and Chen, Yushen and Chen, Peining and Zhu, Pengcheng and Yang, Yunting and Zhang, Bowen and Zhao, Jian and Wang, Chunhui and others},
  journal={arXiv preprint arXiv:2509.22167},
  year={2025}
}

@inproceedings{evans2024fast,
  title={Fast timing-conditioned latent audio diffusion},
  author={Evans, Zach and Carr, CJ and Taylor, Josiah and Hawley, Scott H and Pons, Jordi},
  booktitle={Forty-first International Conference on Machine Learning},
  year={2024}
}

@article{wu2025clear,
  title={Clear: Continuous latent autoregressive modeling for high-quality and low-latency speech synthesis},
  author={Wu, Chun Yat and Deng, Jiajun and Li, Guinan and Kong, Qiuqiang and Lui, Simon},
  journal={arXiv preprint arXiv:2508.19098},
  year={2025}
}

@article{ziyin2020neural,
  title={Neural networks fail to learn periodic functions and how to fix it},
  author={Ziyin, Liu and Hartwig, Tilman and Ueda, Masahito},
  journal={Advances in Neural Information Processing Systems},
  volume={33},
  pages={1583--1594},
  year={2020}
}

@inproceedings{gao2023e3tts,
  title={E3 tts: Easy end-to-end diffusion-based text to speech},
  author={Gao, Yuan and Morioka, Nobuyuki and Zhang, Yu and Chen, Nanxin},
  booktitle={2023 IEEE Automatic Speech Recognition and Understanding Workshop (ASRU)},
  pages={1--8},
  year={2023},
  organization={IEEE}
}

@inproceedings{rombach2022ldm,
  title={High-resolution image synthesis with latent diffusion models},
  author={Rombach, Robin and Blattmann, Andreas and Lorenz, Dominik and Esser, Patrick and Ommer, Bj{\"o}rn},
  booktitle={Proceedings of the IEEE/CVF conference on computer vision and pattern recognition},
  pages={10684--10695},
  year={2022}
}

@inproceedings{sohl2015dpm,
  title={Deep unsupervised learning using nonequilibrium thermodynamics},
  author={Sohl-Dickstein, Jascha and Weiss, Eric and Maheswaranathan, Niru and Ganguli, Surya},
  booktitle={International conference on machine learning},
  pages={2256--2265},
  year={2015},
  organization={pmlr}
}

@article{song2020score,
  title={Score-based generative modeling through stochastic differential equations},
  author={Song, Yang and Sohl-Dickstein, Jascha and Kingma, Diederik P and Kumar, Abhishek and Ermon, Stefano and Poole, Ben},
  journal={arXiv preprint arXiv:2011.13456},
  year={2020}
}

@inproceedings{mehta2024matcha,
  title={Matcha-TTS: A fast TTS architecture with conditional flow matching},
  author={Mehta, Shivam and Tu, Ruibo and Beskow, Jonas and Sz{\'e}kely, {\'E}va and Henter, Gustav Eje},
  booktitle={ICASSP 2024-2024 IEEE International Conference on Acoustics, Speech and Signal Processing (ICASSP)},
  pages={11341--11345},
  year={2024},
  organization={IEEE}
}

@article{jeong2021difftts,
  title={Diff-tts: A denoising diffusion model for text-to-speech},
  author={Jeong, Myeonghun and Kim, Hyeongju and Cheon, Sung Jun and Choi, Byoung Jin and Kim, Nam Soo},
  journal={arXiv preprint arXiv:2104.01409},
  year={2021}
}

@article{albergo2025stochastic,
  title={Stochastic interpolants: A unifying framework for flows and diffusions},
  author={Albergo, Michael and Boffi, Nicholas M and Vanden-Eijnden, Eric},
  journal={Journal of Machine Learning Research},
  volume={26},
  number={209},
  pages={1--80},
  year={2025}
}

@misc{torchdiffeq,
	author={Chen, Ricky T. Q.},
	title={torchdiffeq},
	year={2018},
	url={https://github.com/rtqichen/torchdiffeq},
}

@article{chung2023unimax,
  title={Unimax: Fairer and more effective language sampling for large-scale multilingual pretraining},
  author={Chung, Hyung Won and Constant, Noah and Garcia, Xavier and Roberts, Adam and Tay, Yi and Narang, Sharan and Firat, Orhan},
  journal={arXiv preprint arXiv:2304.09151},
  year={2023}
}

@inproceedings{henry2020qknorm,
  title={Query-key normalization for transformers},
  author={Henry, Alex and Dachapally, Prudhvi Raj and Pawar, Shubham Shantaram and Chen, Yuxuan},
  booktitle={Findings of the Association for Computational Linguistics: EMNLP 2020},
  pages={4246--4253},
  year={2020}
}

@inproceedings{chen2024gentron,
  title={Gentron: Diffusion transformers for image and video generation},
  author={Chen, Shoufa and Xu, Mengmeng and Ren, Jiawei and Cong, Yuren and He, Sen and Xie, Yanping and Sinha, Animesh and Luo, Ping and Xiang, Tao and Perez-Rua, Juan-Manuel},
  booktitle={Proceedings of the IEEE/CVF Conference on Computer Vision and Pattern Recognition},
  pages={6441--6451},
  year={2024}
}

@inproceedings{woo2023convnext2,
  title={Convnext v2: Co-designing and scaling convnets with masked autoencoders},
  author={Woo, Sanghyun and Debnath, Shoubhik and Hu, Ronghang and Chen, Xinlei and Liu, Zhuang and Kweon, In So and Xie, Saining},
  booktitle={Proceedings of the IEEE/CVF conference on computer vision and pattern recognition},
  pages={16133--16142},
  year={2023}
}

@article{xue2022byt5,
  title={ByT5: Towards a token-free future with pre-trained byte-to-byte models},
  author={Xue, Linting and Barua, Aditya and Constant, Noah and Al-Rfou, Rami and Narang, Sharan and Kale, Mihir and Roberts, Adam and Raffel, Colin},
  journal={Transactions of the Association for Computational Linguistics},
  volume={10},
  pages={291--306},
  year={2022},
  publisher={MIT Press One Broadway, 12th Floor, Cambridge, Massachusetts 02142, USA~…}
}

@article{yu2024repa,
  title={Representation alignment for generation: Training diffusion transformers is easier than you think},
  author={Yu, Sihyun and Kwak, Sangkyung and Jang, Huiwon and Jeong, Jongheon and Huang, Jonathan and Shin, Jinwoo and Xie, Saining},
  journal={arXiv preprint arXiv:2410.06940},
  year={2024}
}

@article{boito2024mhubert,
  title={mhubert-147: A compact multilingual hubert model},
  author={Boito, Marcely Zanon and Iyer, Vivek and Lagos, Nikolaos and Besacier, Laurent and Calapodescu, Ioan},
  journal={arXiv preprint arXiv:2406.06371},
  year={2024}
}

@article{kynkaanniemi2024cfgig,
  title={Applying guidance in a limited interval improves sample and distribution quality in diffusion models},
  author={Kynk{\"a}{\"a}nniemi, Tuomas and Aittala, Miika and Karras, Tero and Laine, Samuli and Aila, Timo and Lehtinen, Jaakko},
  journal={Advances in Neural Information Processing Systems},
  volume={37},
  pages={122458--122483},
  year={2024}
}

@inproceedings{sadat2024apg,
  title={Eliminating oversaturation and artifacts of high guidance scales in diffusion models},
  author={Sadat, Seyedmorteza and Hilliges, Otmar and Weber, Romann M},
  booktitle={The Thirteenth International Conference on Learning Representations},
  year={2024}
}

@inproceedings{rix2001pesq,
  title={Perceptual evaluation of speech quality (PESQ)-a new method for speech quality assessment of telephone networks and codecs},
  author={Rix, Antony W and Beerends, John G and Hollier, Michael P and Hekstra, Andries P},
  booktitle={Proc. ICASSP},
  volume={2},
  pages={749--752},
  year={2001},
  organization={IEEE}
}

@article{taal2011stoi,
  title={An algorithm for intelligibility prediction of time--frequency weighted noisy speech},
  author={Taal, Cees H and Hendriks, Richard C and Heusdens, Richard and Jensen, Jesper},
  journal={IEEE Transactions on audio, speech, and language processing},
  volume={19},
  number={7},
  pages={2125--2136},
  year={2011},
  publisher={IEEE}
}

@article{zhou2025voxcpm,
  title={Voxcpm: Tokenizer-free TTS for context-aware speech generation and true-to-life voice cloning},
  author={Zhou, Yixuan and Zeng, Guoyang and Liu, Xin and Li, Xiang and Yu, Renjie and Wang, Ziyang and Ye, Runchuan and Sun, Weiyue and Gui, Jiancheng and Li, Kehan and others},
  journal={arXiv preprint arXiv:2509.24650},
  year={2025}
}

@article{jia2025ditar,
  title={Ditar: Diffusion transformer autoregressive modeling for speech generation},
  author={Jia, Dongya and Chen, Zhuo and Chen, Jiawei and Du, Chenpeng and Wu, Jian and Cong, Jian and Zhuang, Xiaobin and Li, Chumin and Wei, Zhen and Wang, Yuping and others},
  journal={arXiv preprint arXiv:2502.03930},
  year={2025}
}

@article{gong2025acestep,
      title={ACE-Step: A Step Towards Music Generation Foundation Model}, 
      author={Junmin Gong and Sean Zhao and Sen Wang and Shengyuan Xu and Joe Guo},
      year={2025},
      journal={arXiv preprint arXiv:2506.00045}, 
}

@article{xin2024bigcodec,
  title={BigCodec: Pushing the Limits of Low-Bitrate Neural Speech Codec},
  author={Xin, Detai and Tan, Xu and Takamichi, Shinnosuke and Saruwatari, Hiroshi},
  journal={arXiv preprint arXiv:2409.05377},
  year={2024}
}

@article{team2026moss,
  title={MOSS-TTS Technical Report},
  author={SII-OpenMOSS},
  journal={arXiv preprint arXiv:2603.18090},
  year={2026}
}



\end{document}